\documentclass[acus]{JAC2000}
\usepackage{graphicx}
\begin{document}

\title{\flushright{PSN: TUAP003}\\[15pt] \centering Present
        Status of VEPP-5 Control System}

\author{ D.Yu.Bolkhovityanov, R.G.Gromov, E.A.Gousev,
          K.V.Gubin, I.L.Pivovarov, O.Yu.Tokarev\\
	  BINP, Novosibirsk, Russia}

\maketitle

\begin{abstract}
This report concerns the present status of VEPP-5 control system. The control
system hardware consists of CAMAC blocks, a set of crate controllers based on
INMOS transputers and ICL-1900 architecture processor Odrenok, and
Pentium-based workstations. For small tasks simple serial CAMAC controllers
are used. For slow controls of power supplies the CANBUS is begun being used.
The workstations are running Linux and are connected via local net using
TCP/IP. Odrenok crate controllers are joined into other local net and are used
for control of equipment in high voltage pulse condition (klystron gallery).
Transputer crate controllers are linked directly to the server computer and
are used for high performance diagnostics (BPM). The three-level software
complies the so-called ``standard model''.
%
\end{abstract}

\section{Introduction}

\subsection{General design}

\begin{figure}[htbp]
\centering
\includegraphics*[width=\linewidth]{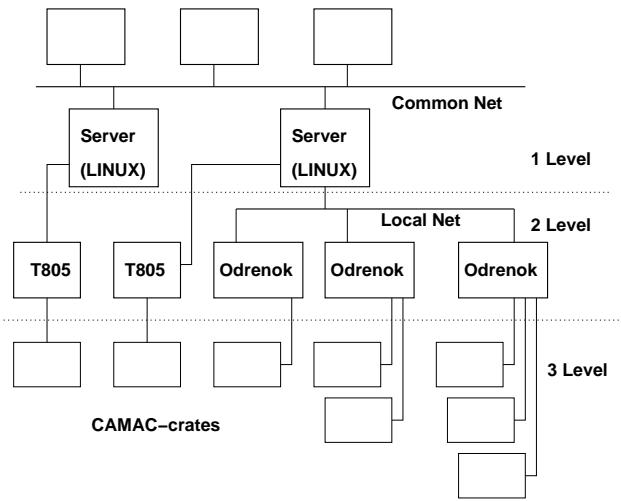}
\caption{Control system network.}
\label{soft:struct}
\end{figure}

VEPP-5 forinjector is a large installation that includes: DC-gun, klystron
gallery, power supply system, bunch compression system, thermo system, BPM and
others \cite{test}. Operation conditions  is pulsed with repetition rate from
1 to 50 Hz.  The control system of VEPP-5 injection complex has a standard
three-level model\cite{gotz} (fig.~\ref{soft:struct}). 

\begin{figure}[htbp]
\centering
\includegraphics*[width=\linewidth]{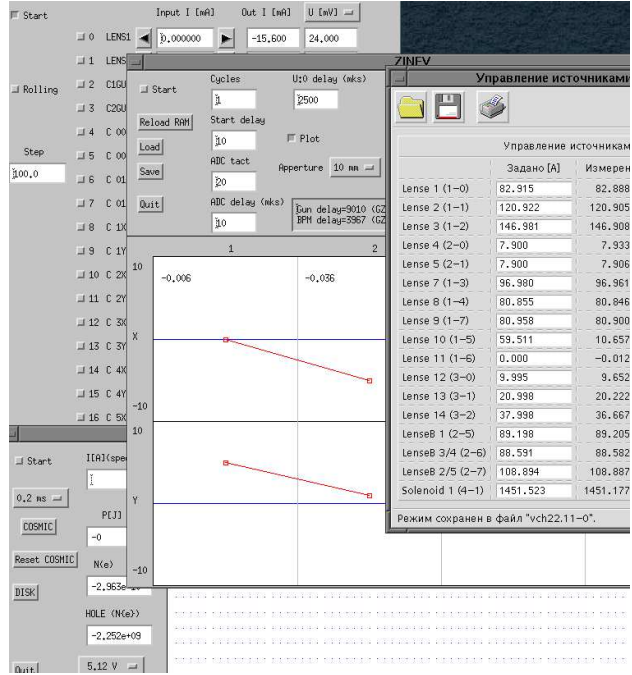}
\caption{Several generations of high-level control software.}
\label{soft:screen}
\end{figure}

The lowest level is composed mainly from CAMAC electronics. Second level is a
set of CAMAC crate controllers with supplementary  low-level software. There
are several types of intellectual crate controllers based on INMOS T805
transputer, ICL-1900 Odrenok and Motorola 5200. 

The high level is the Server which runs under Linux (RedHat-7.1). All client
programs have access to the  low level only via the Server, which performs
initial loading and initialization of the crate controllers and supplementary
programs. 

\subsection{Software design}

Historically there have been several generations of control system software on
VEPP-5.  The very first programs were simple -- they implemented both client
interface and hardware access.  This approach is still used in some
tasks\footnote{A russian proverb says: there's nothing more constant than
temporary.}.

\begin{figure}[htbp]
\centering
\includegraphics*[width=\linewidth]{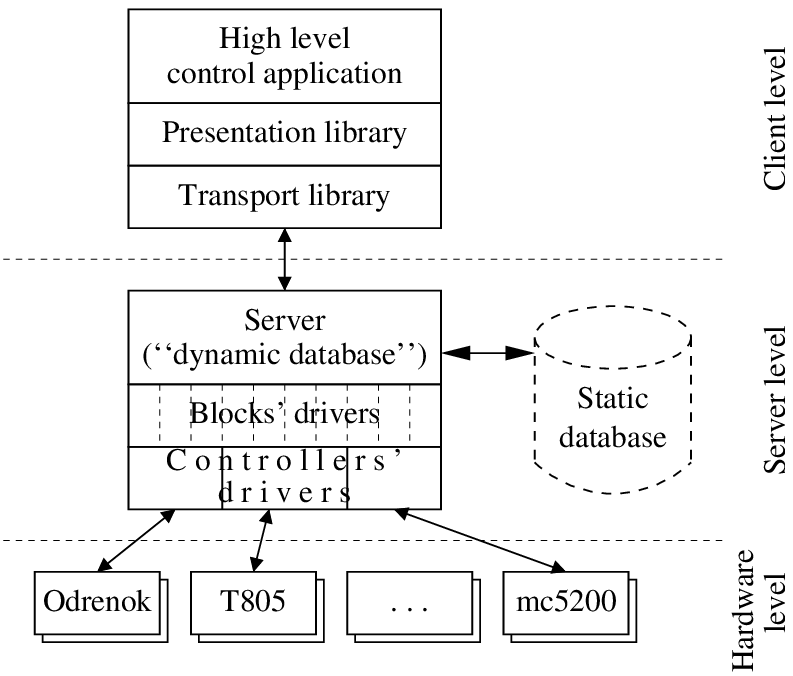}
\caption{Structure of the current
clients~$\leftrightarrow$ server~$\leftrightarrow$ hardware system.}
\label{soft:3levels}
\end{figure}

However, we quickly switched to three-level architecture, as to much more
suitable.  First versions were tied to specific controllers -- separate
software for Odrenok and T805.  They are still in use, but a unified version
was developed which is able to serve various types of controllers
simultaneously (see Fig.\ref{soft:3levels}).  The design of the unified
version was greatly influenced by design principles of X11.

The new version uses the same mechanisms (which are dictated by the nature of
controllers) as previous versions, described in sections \ref{comm:odrenok}
and \ref{comm:t805}.

High-level programs use Motif for client interface.  A special library was
designed, which builds control windows ``on the fly'' from a database
description, thus significantly reducing application creation time \& costs.
(An example of such window can be found at the right of
Fig.\ref{soft:screen}.)

\section{Odrenok $\leftrightarrow$ Server communication}
\label{comm:odrenok}

Odrenok is the most popular crate controller in BINP. It has ICL-1900
instruction set supplemented with commands for CAMAC bus access and vector
operations. New CAMAC adapter was developed to have a possibility to connect
CAMAC with PC via Ethernet. CAMAC-Ethernet adapter provides the speed of data
exchange up to 400kB/sec~\cite{klystron}. It is  sufficient for real-time
network operation. The server workstation has two Ethernet cards: one is for
communication to institute network and another is for local Odrenok net.

\begin{figure}[htbp]
\centering
\includegraphics*[width=70mm]{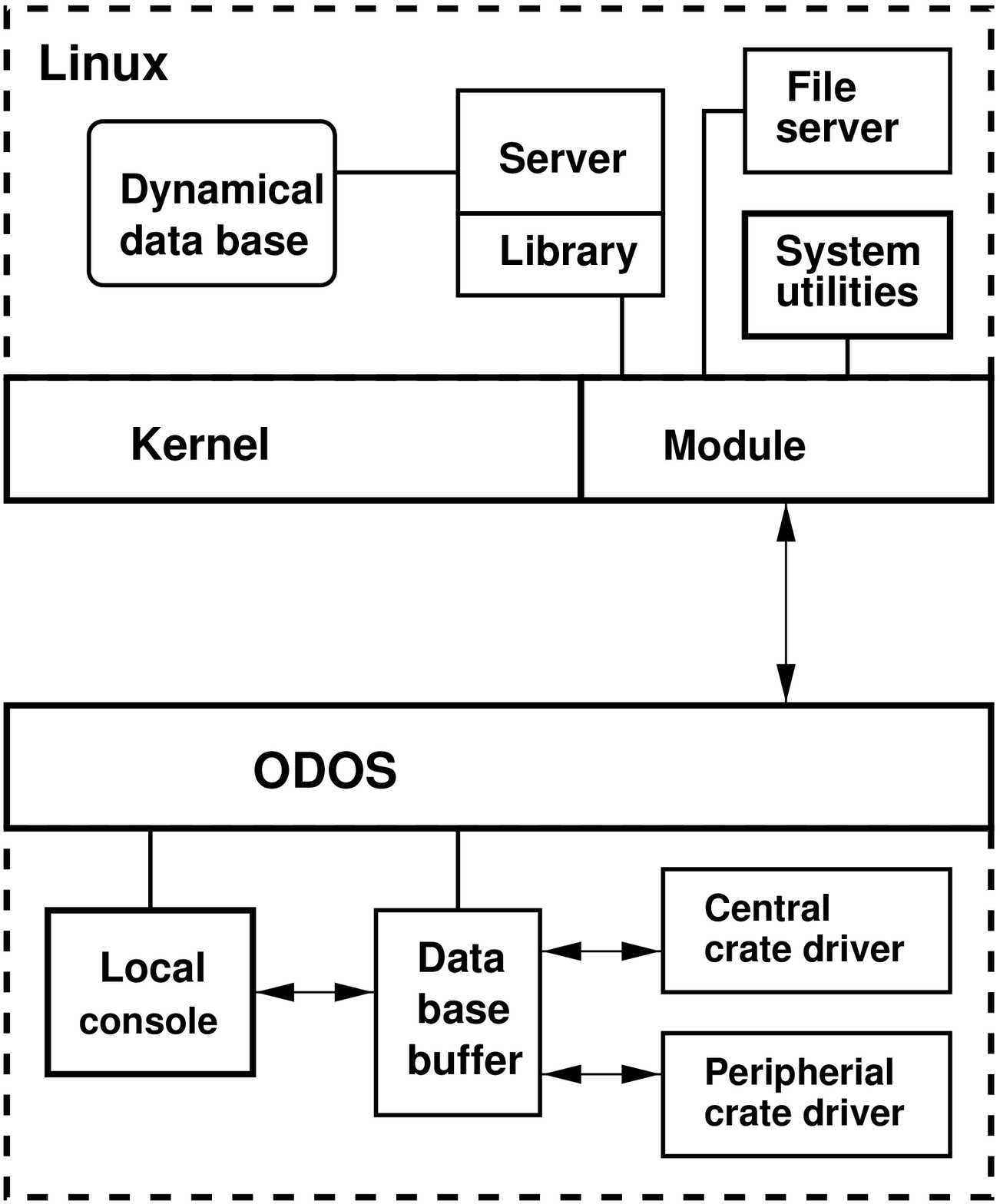}
\caption{Linux $\leftrightarrow$ ODOS Software structure.}
\label{odrenok}
\end{figure}

ODOS (ODrenok Operation System) protocols use Ethernet packets of non-standard
type,  therefore the Server requires I/O facilities for these packets. It is
implemented by kernel module which provides sending, receiving and waiting
functions via standard socket interface; the module is supplemented by client
libraries and utilities (Fig.~\ref{odrenok}).

However, usage of home-made protocol have shown many inconveniences.  So, this
year UDP support was added to ODOS, which enables to use standard Unix
communication interface and solves many other problems (routing between
networks, kernel upgrades, stability).

\section{Transputer $\leftrightarrow$ Server communication}
\label{comm:t805}

Crate controller based on INMOS T805 Transputer~\cite{t805} has been developed
in BINP as high performance device for data asquistion system. This type of
controllers is a suitable tool for wired BPM and pick-up electronic control. 
This is because T805 has a powerful floating point unit.  In this case all
calculations are performed on the 2nd Level and ready data is sent into 1st
Level software.

\begin{figure}[htbp]
\centering
\includegraphics*[width=70mm]{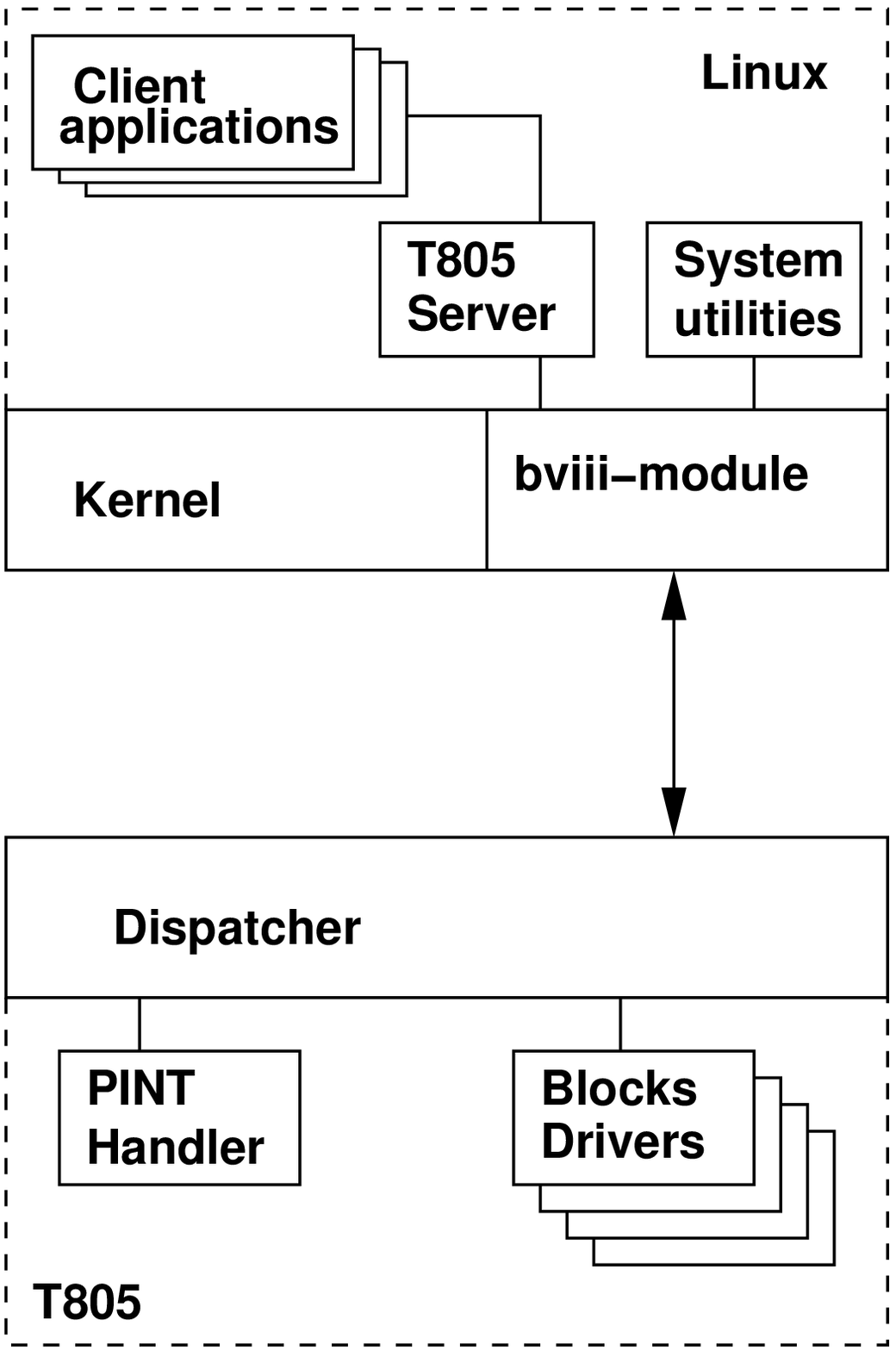}
\caption{Linux $\leftrightarrow$ T805 Software structure.}
\label{inmos}
\end{figure}

Transputer controllers have no OS (despite the fact that transputers are
excellent for parallel tasks), so the following set programs had to be made
(See Fig.~\ref{inmos}).  First, a Dispatcher to perform communication between
local transputer net and a set of block drivers.  Second, an event handler
(``PINT'') which enables drivers to read data when it is ready. Finally, a set
of drivers for individual CAMAC blocks (1 driver process per block).


\section{Motorola crate controller}

Odrenok crate controller is very old.  Production of T805-based controllers
had ceased.  So, we have to find another crate controller.

This year we began using another one, based on Motorola 5200 processor.  100MB
Ethernet is used for comminication to host computers.

This controller runs uClinux~\cite{uclinux} -- a Linux clone designed for
processors without a Memory Management Unit (MMU).  Having Linux in both
workstations and crate controllers significantly eases the life.  On the other
hand, lack of MMU makes multitasking and multithreading too tricky.

So, using ARM, PowerPC or an x86 clone in CAMAC controller could be much
better choice, but 5200 was chosen mainly because of abilities of BINP
electronic design department.

\section{Future development}

Currently the information about hardware structure and knowledge of how it is
mapped to ``physics'' information is spread along the various parts of
software -- in the server config files, hardcoded into application programs,
etc.  This is extremely inconvenient and error-prone, so we switched a
significant part of manpower to design a database, which will contain all this
information (a so-called ``static database'').  The database is based on
PostgreSQL, but for more flexibility all pieces of control software will acces
it through the server.

In the last several years VEPP-5 began to use other hardware in addition to
CAMAC.  In this process we tried to employ as standard interfaces as possible.
The ultimate goal is to replace all custom-made PC$\leftrightarrow$hardware
communication boards with standard ones, such as Ethernet.

For slow controls of power supplies CANBUS devices are used.  However, this
decision is still half-CAMAC-based -- the CANBUS controller is itself a CAMAC
block.  This was done because it was impossible to find PCI CANBUS controllers
with open specifications.  The CANBUS hardware fits nicely into our software
architecture, so we'll widen its use.

Some devices (like TV cameras) require very high bandwidth, which can't be
obtained from CAMAC.  So, our lab designs TV camera with 100MB Ethernet
interface.  Having negative experience with non-standard communication
protocols, we decided to implement transport protocol over UDP.

Since Ethernet chipsets are very cheap, even not-so-demanding hardware is
being implemented with Ethernet as communication media.  Thus Ethernet extends
its presence as one of control system's low-level buses.  But various types of
hardware (crate controllers, TV cameras, slow devices) will be physically
connected to separate buses.

Currently the VEPP-5 control system is being designed and tested on the
forinjector, and in the future it will also be used on damping ring.

\end{document}